\documentclass[aps,prl,twocolumn,showpacs,superscriptaddress,nofootinbib,floatfix]{revtex4}

\usepackage{graphicx}
\usepackage{epsfig}
\usepackage{amsmath}

\newcommand{\be}{\begin{eqnarray}}
\newcommand{\ee}{\end{eqnarray}}
\newcommand{\bea}{\begin{eqnarray}}
\newcommand{\eea}{\end{eqnarray}}

\newcommand{\bra}[1]{\mbox{$\langle #1 |$}}
\newcommand{\ket}[1]{\mbox{$| #1 \rangle$}}
\newcommand{\braket}[2]{\mbox{$\langle #1  | #2 \rangle$}}
\newcommand{\proj}[1]{\mbox{$|#1\rangle \!\langle #1 |$}}

\def\ch{\raisebox{0.3ex}{$\chi$}}
\def\H{{\cal H}}
\def\K{{\cal K}}
\def\J{{\cal J}}

\def\O{{\cal O}}

\DeclareMathOperator{\tr}{tr}

\begin{document}

%%%%%%%%%%%%%%%%%%%%%%%%%%%%%%%%%%%%%%%%%%%%%%%%%%%%%%%%%%%%%%%%%%%%%%%%%%%%%%%

\title{
Efficient classical simulation of \\
slightly entangled 
quantum computations
}

\author{Guifr\'e Vidal}
\affiliation{Institute for Quantum Information, California Institute of
             Technology, Pasadena, CA 91125, USA}

\date{\today}

\begin{abstract}
We present a scheme to efficiently simulate, with a classical computer, the dynamics of multipartite quantum systems on which the amount of entanglement (or of correlations in the case of mixed-state dynamics) is conveniently restricted. The evolution of a pure state of $n$ qubits can be simulated by using computational resources that grow linearly in $n$ and exponentially in the entanglement. We show that a pure-state quantum computation can only yield an exponential speed-up with respect to classical computations if the entanglement increases with the size $n$ of the computation, and gives a lower bound on the required growth.

\end{abstract}


\maketitle

%%%%%%%%%%%%%%%%%%%%%%%%%%%%%%%%%%%%%%%%%%%%%%%%%%%%%%%%%%%%%%%%%%%%%%%%%%%%%%%

In quantum computation, the evolution of a multipartite quantum system is used to efficiently perform computational tasks that are believed to be intractable 
with a classical computer. For instance, provided a series of severe technological difficulties are overcome, Shor's quantum algorithm \cite{Shor} can be used to decompose a large number into its prime factors efficiently ---that is, exponentially faster than with any known classical algorithm.

While it is not yet clear what physical resources are responsible for such suspected quantum computational speed-ups, a central observation, as discussed by Feynman \cite{Feynman}, is that simulating quantum systems by classical means appears to be hard. 
Suppose we want to simulate the joint evolution of $n$ interacting spin systems, each one described by a two-dimensional Hilbert space $\H_2$. Expressing the most general pure state $\ket{\Psi}\in {\H_2}^{\otimes n}$ of the $n$ spins already requires specifying about $2^n$ complex numbers $c_{i_1 \cdots i_n}$,
\be
\ket{\Psi} = \sum_{i_1=0}^1\!\cdots \!\sum_{i_n=0}^1 c_{i_1 \cdots i_n} ~\ket{i_1}\otimes \cdots \otimes \ket{i_n},
\label{eq:compdeco}
\ee
where $\{\ket{0},\ket{1} \in \H_2\}$ denotes a single-spin orthonormal basis; and computing its evolution in time is not any simpler.
This exponential overhead of classical computational resources --as compared to the quantum resources needed to directly implement the physical evolution by using $n$ spin systems-- strongly suggests that quantum systems are indeed computationally more powerful than classical ones.

On the other hand, some specific quantum evolutions can be efficiently simulated by a classical computer -- and therefore cannot yield an exponential computational speed-up. Examples include a system of fermions with only quadratic interactions \cite{fermions}, or a set of two-level systems or qubits initially prepared in a computational-basis state and acted upon by gates from the Clifford group \cite{Clifford}. Recently, Jozsa and Linden \cite{JoLin} have also shown how to efficiently simulate any quantum evolution of an $n$-qubit system when its state factors, at all times, into a product of states each one involving, at most, a constant (i.e. independent of $n$) number of qubits. 

Here we show how to efficiently simulate, with a classical computer, pure-state quantum dynamics of $n$ entangled qubits, whenever only a restricted amount of entanglement is present in the system. It follows that entanglement is a necessary resource in (pure-state) quantum computational speed-ups. More generally, we establish an upper bound, in terms of the amount of entanglement, for the maximal speed-up a quantum computation can achieve. An analogous upper bound, but in terms of correlations (either classical or quantum), also applies to quantum computations with mixed states.

For simplicity sake the analysis is focused on a computation in the quantum circuit model. Thus we consider a discretized evolution of the $n$ qubits, initially in state $\ket{0}^{\otimes n}$, according to a sequence of poly($n$) (i.e., a number polynomial in $n$) single-qubit and two-qubit gates. We recall, however, that {\em any} evolution of $n$ qubits according to single-qubit and two-qubit Hamiltonians can be efficiently approximated, with arbitrary accuracy, by the above circuit model, so that the present results also apply to this more general setting \cite{generalization}.

Consider, as in Eq. (\ref{eq:compdeco}), a pure state $\ket{\Psi} \in {\H_2}^{\otimes n}$ of an $n$-qubit system. Let $A$ denote a subset of the $n$ qubits and $B$ the rest of them. The Schmidt decomposition SD of $\ket{\Psi}$ with respect to the partition $A$:$B$ reads
\be
\ket{\Psi} = \sum_{\alpha=1}^{\ch_A} \lambda_{\alpha}\ket{\Phi_\alpha^{[A]}} \otimes \ket{\Phi_\alpha^{[B]}},
\label{eq:Schmidt}
\ee
where the vector $\ket{\Phi_\alpha^{[A]}}$ ($\ket{\Phi_\alpha^{[B]}}$) is an eigenvector with eigenvalue $|\lambda_\alpha|^{2}>0$ of the reduced density matrix $\rho^{[A]}$ ($\rho^{[B]}$),
whereas the coefficient $\lambda_\alpha$ follows from the relation $\braket{\Phi_\alpha^{[A]}}{\Psi} = \lambda_{\alpha}\ket{\Phi_\alpha^{[B]}}$. The Schmidt rank $\ch_A$ is a natural measure of the entanglement between the qubits in $A$ and those in $B$ \cite{rankmeasure}. Accordingly, we quantify the entanglement of state $\ket{\Psi}$ by $\ch$,
\be
\ch \equiv \max_A \ch_A,
\ee
that is, by the maximal Schmidt rank over all possible bipartite splittings $A$:$B$ of the $n$ qubits. We shall say that $\ket{\Psi}$ is only slightly entangled if $\ch$ is ``small''. In particular, here we are interested in sequences of states $\{\ket{\Psi_n}\}$ of an increasing number $n$ of qubits (corresponding, say, to quantum computations with increasingly large inputs). In such a context we consider $\ch$ to be ``small'' if it grows at most polynomially with $n$,  $\ch_n =$ poly($n$) \cite{maxim}.

\vspace{2mm}

{\bf Definition.--} A pure-state quantum evolution is {\em slightly entangled} if, at all times $t$, the state $\ket{\Psi(t)}$ of the system is slightly entangled ---that is, if $\ch(t)$ is small. A sequence of evolutions with an increasingly large number $n$ of qubits is {\em slightly entangled} if $\ch_n(t)$ is upper bounded by poly$(n)$.

\vspace{2mm}

The key ingredient of our simulation protocol is a {\em local decomposition} of the state $\ket{\Psi} \in {H_2}^{\otimes n}$ in terms of $n$ tensors $\{\Gamma^{[l]}\}_{l=1}^n$ and $n\!-\!1$ vectors $\{\lambda^{[l]}\}_{l=1}^{n\!-\!1}$, denoted
\be
\ket{\Psi} ~~\longleftrightarrow ~~\Gamma^{[1]}\lambda^{[1]}\Gamma^{[2]}\lambda^{[2]} \cdots \Gamma^{[l]} \cdots \lambda^{[n\!-\!1]}\Gamma^{[n]}.
\label{eq:simple2}
\ee
Here, tensor $\Gamma^{[l]}$ is assigned to qubit $l$ and has (at most) three indices, $\Gamma^{[l]i}_{\alpha\alpha'}$, where $\alpha,\alpha' = 1,\cdots, \ch$ and $i=0,1$, whereas $\lambda^{[l]}$ is a vector whose components $\lambda^{[l]}_{\alpha'}$ store the Schmidt coefficients of the splitting $[1\cdots l]$:$[(l\!+\!1)\cdots n]$. More explicitly, we have \cite{product}
\bea
c_{i_1i_2 \cdots i_n} = \!\!\! \sum_{\alpha_1,\cdots,\alpha_{n\!-\!1}}\!\!\! \Gamma_{\alpha_1}^{[1]i_1}  \lambda^{[l]}_{\alpha_1} \Gamma_{\alpha_1 \alpha_2}^{[2]i_2}  \lambda^{[2]}_{\alpha_2} \cdots  \Gamma_{\alpha_{n\!-\!1}}^{[n]i_n}.
\label{eq:superdeco}
\eea
so that the $2^n$ coefficients $c_{i_1\cdots i_n}$ are expressed in terms of about $(2\ch^2+\ch)n$ parameters, a number that grows only linearly in $n$ for a fixed value of $\ch$. This decomposition is local in that, as we shall see, when a two-qubit gate is applied to qubits $l$ and $l\!+\!1$, only $\Gamma^{[l]}$, $\lambda^{[l]}$ and $\Gamma^{[l\!+\!1]}$ need be updated.

Decomposition (\ref{eq:simple2}) (but not $\ch$) depends on the particular way qubits have been ordered from $1$ to $n$, and essentially consists of a concatenation of $n\!-\!1$ SDs. We first compute the SD of $\ket{\Psi}$ according to the bipartite splitting of the systems into qubit 1 and the $n\!-\!1$ remaining qubits [from now on we omit the tensor product symbol],
\bea
\ket{\Psi} &=& \sum_{\alpha_1} \lambda_{\alpha_1}^{[1]} \ket{\Phi_{\alpha_1}^{[1]}}\ket{\Phi^{[2\cdots n]}_{\alpha_1}}\\
&=& \sum_{i_1, \alpha_1} {\Gamma}^{[1]i_1}_{\alpha_1} \lambda_{\alpha_1}^{[1]} \ket{i_1}\ket{\Phi^{[2\cdots n]}_{\alpha_1}}, 
\label{eq:step0}
\eea
where in the last line we have expanded each Schmidt vector $\ket{\Phi_{\alpha_1}^{[1]}} = \sum_{i_1}{\Gamma}^{[1]i_1}_{\alpha_1} \ket{i_1}$ in terms of the basis vectors $\{\ket{0},\ket{1}\}$ for qubit 1. We then proceed according to the following three steps: ($i$) first we expand each Schmidt vector $\ket{\Phi^{[2\cdots n]}_\alpha}$ in a local basis for qubit 2, 
\be
\ket{\Phi^{[2\cdots n]}_{\alpha_1}} = \sum_{i_2} \ket{i_2}\ket{\tau_{\alpha_1i_2}^{[3\cdots n]}};
\label{eq:step1}
\ee
($ii$) then we write each (possibly unnormalized) vector $\ket{\tau_{\alpha_1i_2}^{[3\cdots n]}}$ in terms of the {\em at most $\ch$} Schmidt vectors $\{\ket{\Phi^{[3\cdots n]}_{\alpha_2}}\}_{\alpha_2=1}^{\ch}$ (i.e., the eigenvectors of $\rho^{[3\cdots n]}$) and the corresponding Schmidt coefficients $\lambda_{\alpha_2}^{[2]}$, 
\be
\ket{\tau_{\alpha_1i_2}^{[3\cdots n]}} = \sum_{\alpha_2} {\Gamma}^{[2]i_2}_{\alpha_1\alpha_2} \lambda^{[2]}_{\alpha_2}\ket{\Phi^{[3\cdots n]}_{\alpha_2}};
\label{eq:step2}
\ee
($iii$) finally we substitute Eq. (\ref{eq:step2}) in Eq. (\ref{eq:step1}) and the latter in Eq. (\ref{eq:step0}) to obtain
\be
\ket{\Psi} = \sum_{i_1,\alpha_1,i_2,\alpha_2} {\Gamma}^{[1]i_1}_{\alpha_1} \lambda_{\alpha_1}^{[1]} {\Gamma}^{[2]i_2}_{\alpha_1\alpha_2} \lambda^{[2]}_{\alpha_2}\ket{i_1i_2}\ket{\Phi^{[3\cdots n]}_{\alpha_1}}.
\ee

Iterating steps ($i$)-($iii$) for the Schmidt vectors $\ket{\Phi_{\alpha_2}^{[3\cdots n]}}$, $\ket{\Phi_{\alpha_3}^{[4\cdots n]}}$, $\cdots$, $\ket{\Phi_{\alpha_{n\!-\!1}}^{[n]}}$, one can express state $\ket{\Psi}$ in terms of tensors ${\Gamma}^{[l]}$ and $\lambda^{[l]}$, as in Eq. (\ref{eq:simple2}).

A useful feature of description (\ref{eq:simple2}) is that it readily gives the SD of $\ket{\Psi}$ according to the bipartite splitting $[1\cdots l]:[(l\!+\!1)\cdots n]$,
\be
\ket{\Psi} = \sum_{\alpha_l} \lambda^{[l]}_{\alpha_l} \ket{\Phi^{[1\cdots l]}_{\alpha_l}} \ket{\Phi^{[(l\!+\!1)\cdots n]}_{\alpha_l}}.
\ee
Indeed, it can be checked by induction over $l$ that 
\be
\ket{\Phi^{[1\cdots l]}_{\alpha_l}} \longleftrightarrow 
{\Gamma}^{[1]}\lambda^{[1]} \cdots \lambda^{[l\!-\!1]} {\Gamma}^{[l]}_{\alpha_l},
\ee
meaning that
\be
\ket{\Phi^{[1\cdots l]}_{\alpha_l}} = \!\!\!\!\!
\sum_{\alpha_1,\cdots,\alpha_{l\!-\!1}}\!\!\!\!\! 
{\Gamma}_{\alpha_1}^{[1]i_1} \lambda^{[1]}_{\alpha_1} \cdots  {\Gamma}_{\alpha_{l\!-\!1} \alpha_l}^{[l]i_l} \ket{i_1\cdots i_l}; 
\ee 
whereas by construction we already had that
\be
\ket{\Phi^{[(l\!+\!1)\cdots n]}_{\alpha_l}} \longleftrightarrow 
{\Gamma}^{[l+1]}_{\alpha_l}\lambda^{[l+1]} \cdots \lambda^{[n\!-\!1]} {\Gamma}^{[n]},
\ee
which stands for
\be
\ket{\Phi^{[(l\!+\!1)\cdots n]}_{\alpha_l}} = \!\! \!\!\!\!\!
\sum_{\alpha_{l\!+\!1},\cdots,\alpha_{n}}\!\!\!\!\! 
{\Gamma}_{\alpha_{l}\alpha_{l\!+\!1}}^{[l\!+\!1]i_{l\!+\!1}} \cdots \lambda^{[n\!-\!1]}_{\alpha_{n\!-\!1}} {\Gamma}_{\alpha_{n\!-\!1}}^{[n]i_n} \ket{i_{l\!+\!1}\cdots i_n}.
\ee

The following lemmas explain how to update the description of state $\ket{\Psi}$ when a single-qubit gate or a two-qubit gate (acting on consecutive qubits) is applied to the system. Remarkably, the computational cost of the updating is independent of the number $n$ of qubits, and only grows in $\ch$ as a polynomial of low degree.

\vspace{2mm}

{\bf Lemma 1.--} Updating the description (\ref{eq:simple2}) of state $\ket{\Psi}$ after a unitary operation $U$ acts on qubit $l$ does only involve transforming ${\Gamma}^{[l]}$. The incurred computational cost is of $\O(\ch^2)$ basic operations.

\vspace{2mm}

{\bf Proof.--} In the SD according to the splitting $[1\cdots (l\!-\!1)]:[l\cdots n]$, a unitary operation $U$ on qubit $l$ does not modify the Schmidt vectors for part $[1\cdots (l\!-\!1)]$ and therefore ${\Gamma}^{[j]}$ and $\lambda^{[j]}$ $(1\leq j \leq l\!-\!1)$ remain the same. Similarly, by considering the SD for the splitting $[1\cdots l]:[(l\!+\!1)\cdots n]$, we conclude that also ${\Gamma}^{[j]}$ and $\lambda^{[j-1]}$ $(l\!+\!1 \leq j \leq n)$ remain unaffected. Instead, ${\Gamma}^{[l]}$ changes according to
\be
{\Gamma'}^{[l]{i}}_{\alpha\beta} =  \sum_{j=0,1}U_{j}^i{\Gamma}^{[l]j}_{\alpha\beta}~~~~~~ \forall \alpha, \beta = 1, \cdots, \ch.
\ee

\vspace{2mm}

{\bf Lemma 2.--} Updating the description (\ref{eq:simple2}) of state $\ket{\Psi}$ after a unitary operation $V$ acts on qubits $l$ and $l+1$ does only involve transforming ${\Gamma}^{[l]}$, $\lambda^{[l]}$ and ${\Gamma}^{[l\!+\!1]}$. This can be achieved with $\O(\ch^3)$ basic operations.

\vspace{2mm}

{\bf Proof.--} In order to ease the notation we regard $\ket{\Psi}$ as belonging to only $4$ subsystems,
\be
\H = \J \otimes \H_C\otimes \H_D\otimes \K.
\ee
Here, $\J$ is spanned by the $\ch$ eigenvectors of the reduced density matrix
\be
\rho^{[1\cdots (l\!-\!1)]}= \sum_{\alpha} \proj{\alpha},  ~~~ \ket{\alpha} \equiv \lambda^{[l-1]}_{\alpha} \ket{\Phi^{[1\cdots (l\!-\!1)]}_{\alpha}};
\ee
and, similarly, $\K$ is spanned by the $\ch$ eigenvectors of the reduced density matrix 
\be
\rho^{[(l\!+\!2) \cdots n]}= \sum_{\gamma} \proj{\gamma}, ~~~\ket{\gamma} \equiv \lambda^{[l+1]}_{\gamma} \ket{\Phi^{[(l\!+\!2)\cdots n]}_{\gamma}};
\ee
whereas $\H_C$ and $\H_D$ correspond, respectively, to qubits $l$ and $l\!+\!1$. In this notation we have
\be
\ket{\Psi} =  \sum_{\alpha, \beta, \gamma=1}^{\ch} \sum_{i,j=0}^1 {\Gamma}^{[C]i}_{\alpha\beta}\lambda_{\beta}{\Gamma}^{[D]j}_{\beta \gamma} \ket{\alpha ij\gamma},
\ee
and, reasoning as in the proof of lemma 1, when applying unitary $V$ to qubits $C$ and $D$ we need only update ${\Gamma}^{[C]}, \lambda, {\Gamma}^{[D]}$. We can expand $\ket{\Psi'} \equiv V\ket{\Psi}$ as
\be
\ket{\Psi'} = \sum_{\alpha, \gamma=1}^{\ch} \sum_{i,j=0}^1 \Theta_{\alpha\gamma}^{ij} \ket{\alpha ij\gamma},
\ee
where
\be
\Theta_{\alpha\gamma}^{ij} = \sum_{\beta}\sum_{kl} V^{ij}_{kl} {\Gamma}^{[C]k}_{\alpha\beta}\lambda_{\beta}{\Gamma}^{[D]l}_{\beta \gamma}.
\ee
By diagonalizing $\rho'^{[D\K]}$,
\bea
\rho'^{[D\K]} &=& \tr_{\J C} \proj{\Psi'} \\ 
&=& \!\!\!\!\!\sum_{j,j', \gamma, \gamma'} \!\!\left( \sum_{\alpha,i} \braket{\alpha}{\alpha}\Theta_{\alpha\gamma}^{ij} (\Theta_{\alpha\gamma'}^{ij'})^*\!\right) \ket{j\gamma}\bra{j'\gamma'}, \nonumber
\eea
we obtain its eigenvectors $\{\ket{\Phi'^{[D\K]}_{\beta}}\}$, which we can expand in terms of $\{\ket{j\gamma}\}$ to obtain ${\Gamma}'^{[D]}$,
\be
\ket{\Phi'^{[D\K]}_{\beta}} = \sum_{j,\gamma} {\Gamma}'^{[D]j}_{\beta\gamma} \ket{j\gamma}.
\label{eq:extract}
\ee
The eigenvectors of $\rho'^{[\J C]}$ and $\lambda'$ follow then from
\bea
\lambda'_\beta \ket{\Phi'^{[\J C]}_{\beta}} &=& \braket{\Phi'^{[D\K]}_\beta}{\Psi'}\\
&=& \sum_{i,j,\alpha,\gamma}  ({\Gamma}'^{[D]j}_{\beta\gamma})^*\Theta^{ij}_{\alpha\gamma} \braket{\gamma}{\gamma} \ket{\alpha i},
\eea
and by expanding each $\ket{\Phi'^{[\J C]}_\beta}$,
\be
\ket{\Phi'^{[\J C]}_\beta} = \sum_{i\alpha} {\Gamma'}^{[C]i}_{\alpha\beta} \ket{\alpha i},
\ee
we also obtain ${\Gamma'}^{[C]}$. All the above manipulations can be performed by storing $\O(\ch^2)$ coefficients and require $\O(\ch^3)$ basic operations.

We now state our main results. We consider a pure-state quantum computation using $n$ qubits, and consisting of poly$(n)$ one- and two-qubit gates and a final local measurement. The simulation protocol works as follows. We use tensors ${\Gamma}^{[l]}$ and $\lambda^{[l]}$ to store the initial state $\ket{0}^{\otimes n}$ and update its description as the gates are applied \cite{fractional}. Recall that in description (\ref{eq:simple2}) each qubit has been associated a position from $1$ to $n$. In order to update $\ket{\Psi}$ according to a two-qubit gate between non-consecutive qubits $C$ and $D$, we will first simulate $\O(n)$ swap gates between adjacent qubits to bring $C$ and $D$ together. Computing the expectation value for any product operator (e.g. a projection corresponding to a local measurement) from $\{\Gamma^{[l]},\lambda^{[l]}\}$ is straightforward and can also be done with $n$ poly($\ch$) operations.

\vspace{2mm}

{\bf Theorem 1.---} If through a pure-state quantum computation $\ch_n$ is upper bounded by poly($n$), then the computation can be classically simulated with poly$(n)$ memory space and computational time.

\vspace{2mm}

{\bf Theorem 2.---} If $\ch_n$ grows subexponentially in $n$, then the quantum computation can be classically simulated with subexp$(n)$ memory space and computational time.

\vspace{2mm}

Thus, theorem 1 provides us with a {\em sufficient} condition for the efficient classical simulation of a quantum computation, which by extension also applies to generic pure-state, multi-particle unitary dynamics generated by local interactions \cite{generalization}. In turn theorem 2 provides us with a more general condition under which a quantum computation cannot yield an exponential speed-up with respect to classical computations. Both theorems follow straightforwardly from the previous lemmas and considerations.

The above results establish a clear connection between the amount of entanglement in a multipartite system and the computational cost of simulating the system with a classical computer. This suggests a new approach to the study of multipartite entanglement, based on the complexity of describing and simulating quantum systems. We propose to quantify the entanglement of a pure state $\ket{\Psi}$ through measures that indicate how difficult it is to express $\ket{\Psi}$ in terms of local states or, relatedly, to account for a local change in the system. An example of such entanglement measures is the function
\be
E_{\ch} \equiv \log_2 \ch,
\ee 
which, apart from serving the purposes, has a series of other appealing properties: ($i$) $E\chi$ only vanishes for product (i.e., unentangled) vectors; ($ii$) $E\chi$ is additive under tensor products, $E_{\ch}(\Psi\otimes\Psi') = E_{\ch}(\Psi)+E_{\ch}(\Psi')$; ($iii$) $E\chi$ monotonically decreases under (both deterministic and stochastic) LOCC manipulations of the system. We also note that $E\chi(\Psi)$ is not a continuous function of $\ket{\Psi}$ with respect any reasonable distance \cite{discontinuous}.

We can rephrase the results of this paper in terms of $E\chi$. Notice that the maximum value of $E_{\ch}$ in a system of $n$ particles is {\em linear} in $n$. Theorem 1 states that an efficient simulation of quantum dynamics is possible whenever $E_{\ch}$ grows at most {\em logarithmically} in $n$. More generally, we have shown how a state $\ket{\Psi}$ can be given a description in terms of local states by using a number of parameters that grows linearly in the number of systems and exponentially in the amount of entanglement $E\chi$,
\be
\begin{array}{cc}\mbox{local description } \\ 
 \mbox{of an } n \mbox{-qubit state} 
\end{array}
\approx
\begin{array}{cc}
 n\exp (E\chi)\\ \mbox{ parameters.}
\end{array}
\ee
This expression implies an upper bound, in terms of the entanglement, for the computational speed-up a quantum evolution can achieve with respect to classical computations.

So far we have only considered pure-state dynamics. But if the $n$ qubits are in a mixed state $\rho \in {\cal B}({\H_2}^{\otimes n})$, we can regard density matrices as vectors in the space of linear operators. By using product expansions and the Schmidt decomposition in this space, one can readily re-derive the above results, but with the former role of entanglement played now by both quantum and classical correlations. Thus, an efficient simulation is possible if the total amount of correlations (as measured by the analog of $\ch$) is sufficiently restricted. In particular, this results do not rule out the possibility of obtaining a computational speed-up through a quantum computation with very noisy mixed states \cite{knill}.

Finally, [a simple modification of] the simulation protocol discussed in this paper may find practical applications as a tool to study quantum systems \cite{prep}. The results of \cite{trans} suggest that, at zero temperature, non-critical spin-chains typically meet sufficient conditions for an efficient classical simulation. Perhaps, then, understanding the structure of multipartite entanglement is the key to achieve efficient simulation of certain multipartite quantum phenomena.

%%%%%%%%%%%%%%%%%%%%%%%%%%%%%%%%%%%%%%%%%%%%%%%%%%%%%%%%%%%%%%%%%%%%%%%%%%%%%%%

\medskip

The author thanks Dave Bacon, Ignacio Cirac, Ann Harvey and Richard Jozsa and Debbie Leung for valuable advice. Support from the US National Science Foundation under Grant No. EIA-0086038 is acknowledged.

%%%%%%%%%%%%%%%%%%%%%%%%%%%%%%%%%%%%%%%%%%%%%%%%%%%%%%%%%%%%%%%%%%%%%%%%%%%%%%%


\begin{thebibliography}{99}
\bibitem{generalization} A straightforward generalization of the present results is possible to the case of $d$-level subsystems with $m$-body interactions, provided $d$ and $m$ do not grow with the total number $n$ of subsystems.
\bibitem{maxim} For a general $n$-qubit state, $\ch_A$ is upper bounded by $2^{n/2}$ [value reached when $A$ contains half of the $n$ qubits and, e.g., $\ket{\Psi}$ is maximally entangled between $A$ and $B$]. Thus, states for which $\ch = $ poly($n$) contain exceptionally little entanglement.
\bibitem{product} Expansion (\ref{eq:superdeco}) very much resembles that of a product vector $\ket{\Psi_{prod}}= \ket{\Phi^{[1]}} \otimes \cdots \otimes \ket{\Phi^{[n]}}$, in which case the coefficients $c_{i_1\cdots i_l \cdots i_n} = \Gamma^{[1]i_1}\cdots \Gamma^{[l]i_l} \cdots \Gamma^{[n]i_n}$ can be expressed in terms of tensors $\Gamma^{[l]}$, where $\Gamma^{[l]}$ completely characterizes the pure state $\ket{\Phi^{[l]}}$ of qubit $l$. The extra indices $\alpha$'s in (\ref{eq:superdeco}) account for the correlations between qubits.
\bibitem{fractional} A digital computer only allows for an approximate description of gates and states, since real coefficients are truncated. See Ref. \cite{JoLin} for a discussion on how to obtain efficient approximations by using rational numbers.
\end{thebibliography}
\end{document}